\documentclass[prl,twocolumn,showpacs,superscriptaddress,psfig]{revtex4}
\usepackage{amssymb}
\usepackage{graphicx}
\usepackage{dcolumn}
\usepackage{bm}

\begin{document}

\title{Evidence of nodal gap structure in the non-centrosymmetric
superconductor Y$_2$C$_3$}
\author{J. Chen}
\affiliation{Department of Physics, Zhejiang University, Hangzhou, Zhejiang 310027, China}
\author{M. B. Salamon}
\affiliation{Department of Physics, University of Texas at Dallas, Richardson, Texas
75080, USA}
\author{S. Akutagawa}
\affiliation{Department of Physics and Mathematics, Aoyama Gakuin University, Sagamihara,
Kanagawa 229-8558, Japan}
\author{J. Akimitsu}
\affiliation{Department of Physics and Mathematics, Aoyama Gakuin University, Sagamihara,
Kanagawa 229-8558, Japan}
\author{J. Singleton}
\affiliation{NHMFL, Los Alamos National Laboratory, MS E536, Los
Alamos, NM 87545, USA}
\author{H. Q. Yuan} \affiliation{Department of
Physics, Zhejiang University, Hangzhou, Zhejiang 310027, China}
\date{\today }

\begin{abstract}
The magnetic penetration depth $\lambda (T)$ and the upper critical field $%
\mu _{0}H_{c2}(T_{c})$ of the non-centrosymmetric (NCS) superconductor Y$_{2}
$C$_{3}$ have been measured using a tunnel-diode (TDO) based resonant
oscillation technique. We found that the penetration depth $\lambda (T)$ and
its corresponding superfluid density $\rho _{s}(T)$ show linear temperature
dependence at very low temperatures ($T\ll T_{c}$), indicating the existence
of line nodes in the superconducting energy gap. Moreover, the upper
critical field $\mu _{0}H_{c2}(T_{c})$ presents an upturn at low
temperatures with a rather high value of $\mu _{0}H_{c2}(0)$ $\simeq 29$T,
which slightly exceeds the weak-coupling Pauli limit. We discuss the
possible origins for these nontrivial superconducting properties, and argue
that the nodal gap structure in Y$_{2}$C$_{3}$ is likely attributed to the
absence of inversion symmetry, which allows the admixture of spin-singlet
and spin-triplet pairing states.
\end{abstract}

\maketitle





The symmetries of a superconductor, e.g., time-reversal and spatial
inversion symmetries, may impose important constraints on the
pairing states. Among the previously investigated superconductors,
most possess an inversion center in their crystal structures, in
which the Cooper pairs are either in an even-parity spin-singlet or
odd-parity spin-triplet pairing state, according to the Pauli
principle and parity conservation \cite{Anderson59, Anderson84}.
However, the tie between spatial symmetry and the Cooper-pair spins
might be violated in a superconductor lacking inversion symmetry
\cite{Gor'kov, Yip,Frigeri,Samokhin04,Fujimoto9}. In these
materials, an asymmetric potential gradient yields an antisymmetric
spin-orbit coupling (ASOC), which splits the Fermi surface into two
spin-ordered subsurfaces, with pairing allowed both across one
subsurface and between the two. ASOC splits the degeneracy of
conduction electron, and allows the admixture of spin-singlet and
spin-triplet pairing states within the same orbital channel. Various
exotic electromagnetic properties may arise in the parity-violated
materials \cite{Gor'kov, Yip,Frigeri,Samokhin04,Fujimoto9}.

Until now, only a very few non-centrosymmetric (NCS) superconductors have
been investigated. In the heavy fermion NCS superconductors, e.g., CePt$_{3}$%
Si and CeMSi$_{3}$ (M=Rh, Ir), the upper critical field is greatly enhanced
in comparison with that of other heavy fermion materials \cite%
{Bauer04,Kimura98,Settai08}, exceeding the Pauli limit in a weak coupling
BCS model. Measurements of thermodynamic properties in CePt$_{3}$Si
indicated the existence of line nodes in the superconducting gap, even
though a weak Hebel-Slichter coherence peak is observed below $T_{c}$ \cite%
{Yogi04,Bonalde,Izawa05,Takeuchi}. These unconventional features were
interpreted in terms of a Rashba-type spin-orbit coupling model arising from
the absence of inversion symmetry \cite{Frigeri,Samokhin04,Fujimoto9}.
However, the strong electronic correlations and the closeness to a magnetic
instability might complicate the analysis of the ASOC effect on
superconductivity. Recent work on Li$_{2}$(Pd$_{1-x}$Pt$_{x}$)$_{3}$B
systems demonstrated that spin-singlet and spin-triplet order parameters can
add constructively and destructively, leading to two different gap functions
and the possible occurrence of accidental nodes in the destructive gap \cite%
{Yuan06,Yuan08,Nishi07,Takeya}. Mo$_{3}$Al$_{2}$C seems to present another
example of weakly correlated NCS superconductor \cite{Bauer10}, showing
evidence of unconventional pairing state attributed to the lack of inversion
symmetry. To look into the effect of ASOC on superconductivity,\ a search
for unconventional superconductivity in weakly correlated
non-centrosymmetric compounds remains highly desireable.

The nonoxide transition metal sequicarbides M$_{2}$C$_{3}$ (M=La, Y), which
crystallize in a cubic Pu$_{2}$C$_{3}$-type structure (space group I$%
\overline{4}3$d), present another family of NCS superconductor with a
relatively high $T_{c}$ ($T_{c}\sim 18$ K)\cite{Krupka, Amano}. Resembling
the Li$_{2}$(Pd$_{1-x}$Pt$_{x}$)$_{3}$B system \cite%
{Yuan06,Yuan08,Nishi07,Takeya} and Mo$_{3}$Al$_{2}$C \cite{Bauer10}, M$_{2}$C%
$_{3}$ shows no evidence of strong electronic correlations and/or magnetic
fluctuations. Thus it might provide another example in which to study the
effect of ASOC on superconductivity. Measurements of
nuclear-magnetic-resonance (NMR) \cite{Harada} and muon spin relaxation ($%
\mu $SR) \cite{Kuroiwa} indicate a complex gap structure in
Y$_{2}$C$_{3}$ which cannot be described in terms of a simple s-wave
or a d-wave gap function, but can be qualitatively fitted with a
two-gap model at temperatures near $T_{c}$. While there is some
variation among the weights reported, the fits near $T_{c}$ all give
a dominant large gap of $2\Delta
_{1}/k_{B}T_{c}$ $\approx 5$ and a small gap of $2\Delta _{2}/k_{B}T_{c}$ $%
\approx$ 1.1-2. On the other hand, the Knight shift in NMR decreases
to approximately $2/3$ of its normal-state value, and the
temperature dependence of $1$/$T_{1}$ deviates from the predictions
of a conventional BCS model \cite{Harada}. Recently, we noticed that
Y$_{2}$C$_{3}$ shows a high upper critical field ($\mu
_{0}H_{c2}(0)$ $\simeq 29$T), which might indicate the importance of
ASOC effect in this compound \cite{Yuan10}.

In order to elucidate further the pairing state in Y$_{2}$C$_{3}$,we report
in this letter an accurate measurement of the temperature dependence of the
resonant frequency shift $\Delta f(T)$ down to 90mK using a TDO-based
resonant oscillator. The penetration depth $\lambda (T)$ and the
corresponding superfluid density $\rho _{s}(T)$ derived from $\Delta f(T)$
can be well interpreted by a two-gap model at temperatures near $T_{c}$.
Remarkably, however, a linear-type temperature dependence of $\lambda (T)$
and $\rho _{s}(T)$ is observed at very low temperatures ($T\ll T_{c}$),
providing evidence for the existence of line nodes in the energy gap.
Together with the observation of a high upper critical field ($\mu
_{0}H_{c2}(0)$ $\simeq 29$T) in Y$_{2}$C$_{3}$, we argue that ASOC might
give rise to the admixture of spin-singlet and spin-triplet pairing states
as previously discussed in Li$_{2}$(Pd$_{1-x}$Pt$_{x}$)$_{3}$B systems \cite%
{Yuan06}, leading to the appearance of a nodal gap structure at low
temperatures.


Polycrystalline samples of Y$_{2}$C$_{3}$ were prepared by an arc-melting
method, followed by heat treatments under high-temperature and high-pressure
conditions to form the sesquicarbide phase \cite{Amano}. Powder X-ray
diffraction identified the achieved ingots as a single phase. Precise
measurements of the resonant frequency shift $\Delta f(T)$ were performed
using a TDO-based, self-inductive technique at $31$MHz in a dilution
refrigerator down to 90 mK \cite{Chia}. The change in penetration depth $%
\Delta \lambda (T)$ is proportional to $\Delta f(T)$, i.e., $\Delta \lambda
(T)=G\cdot \Delta f(T)$, where the G-factor is a constant determined by
sample and coil geometries and, therefore, varies from sample to sample. It
is noted that the samples of Y$_{2}$C$_{3}$ investigated here are rather air
sensitive and we could not polish the surface to form a regular geometry,
which might result in a relatively large uncertainty in the determination of
the G-factor when following traditional methods as described in Ref. \cite%
{Chia}. Nevertheless the temperature-dependent behavior of $\Delta \lambda
(T)$ remains reliable and unchanged. The upper critical field $\mu
_{0}H_{c2}(T_{c})$ was also determined from a similar TDO-based resonant
oscillation method in a pulsed magnetic field \cite{Yuan10}.


\begin{figure}[b]
\centering
\includegraphics[width=7cm]{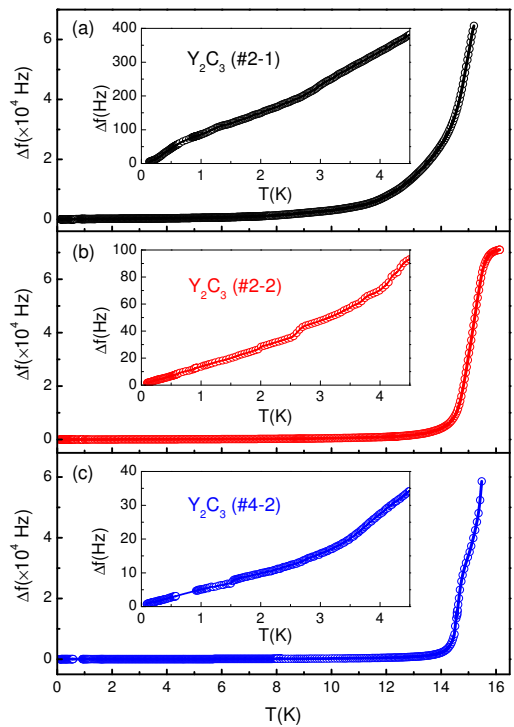}
\caption{(Color online) Temperature dependence of the resonant frequency
shift $\Delta f(T)$ for variant Y$_2$C$_3$ samples. The insets plot $\Delta
f(T)$ at low temperatures, showing linear-type temperature dependence.}
\label{fig1}
\end{figure}

In Fig. 1, we plot the temperature dependence of $\Delta f(T)$ for three
samples of Y$_{2}$C$_{3}$, which were cut either from the same batch (\#2-1,
\#2-2) or from a different batch (\#4-2). One can see that all three samples
follow similar behavior. The sharp drop of $\Delta f(T)$ marks a
superconducting transition with $T_{c}\simeq $15 K. The linear temperature
dependence of $\Delta f(T)$ is reproducibly achieved at low temperatures as
shown in the insets of Fig. 1, significantly deviating from the exponential
behavior of conventional BCS superconductors. Such linear temperature
dependence of $\Delta \lambda (T)$ gives strong evidence for the existence
of line nodes in the superconducting energy gap of Y$_{2}$C$_{3}$,
remarkably resembling the case of Li$_{2}$(Pd$_{1-x}$Pt$_{x}$)$_{3}$B ($%
x\geq 0.3$) \cite{Yuan06,Yuan08}. The different absolute values of $\Delta
f(T)$ are mainly attributed to the distinct values of the G-factor for each
sample.

\begin{figure}[b]
\centering
\includegraphics[width=7cm]{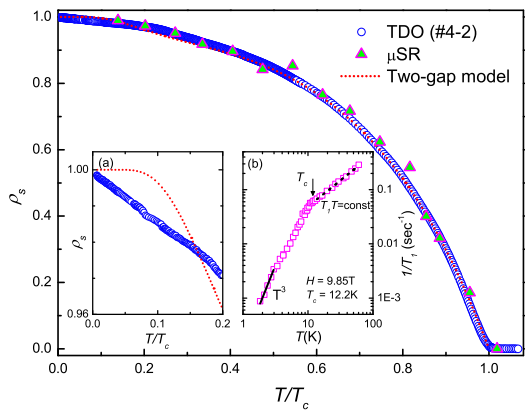}
\caption{(Color online) The normalized superfluid density
$\protect\rho_s(T)$ versus temperature for Y$_2$C$_3$. The circles
represent our experimental data derived from the penetration depth
$\protect\lambda(T)$ of sample \#4-2. The filled triangles are
digitized from Ref. \protect\cite{Kuroiwa} which were determined
from the $\protect\mu$SR experiments. The dotted line indicates the
best fitting using the two-gap model described in the text. Inset
(a) shows $\protect\rho_s(T)$ and the fittings in the low
temperature region and (b) shows the temperature dependence of
$1/T_1$ for Y$_2$C$_3$ (from Ref. \protect\cite{Harada}),
respectively.} \label{fig2}
\end{figure}

The superfluid density is directly related to the superconducting gap
structure on the Fermi surface. In Fig. 2, we plot the superfluid density $%
\rho _{s}(T)$ for Y$_{2}$C$_{3}$ (circles), derived by $\rho _{s}(T)=\lambda
^{2}(0)/\lambda ^{2}(T)$. Here $\lambda (T)=\Delta \lambda (T)+\lambda (0)$,
in which $\lambda (0)=470$ nm is the penetration depth at zero temperature
\cite{Kuroiwa}, and the G-factor is $0.45$ nm/Hz for sample \#4-2. For
comparison, $\rho _{s}(T)$ obtained from the $\mu $SR experiments is also
shown in Fig. 2 (see the triangles in the main figure)\cite{Kuroiwa}. The
agreement between the two measurements is excellent. However, our
penetration depth measurements extend $\rho _{s}(T)$ to much lower
temperatures, which provides significant inputs on the superconducting
pairing state (see below).

A two-gap scenario, as previously discussed in MgB$_2$ \cite{Bouquet,Liu01}%
, has been proposed to describe the NMR and $\mu$SR data for Y$_2$C$_3$ \cite%
{Harada,Kuroiwa}. Here we tried to fit $\rho_s(T)$ with a two-gap BCS model:
$\rho_s(T)=x\delta\rho_s(\Delta_1(0),T)+(1-x)\delta\rho_s(\Delta_2(0),T)$,
where $\Delta_i$ ($i=1$ and $2$) are the energy gaps at $T=0$ and $x$ is the
relative weight for the $\Delta_1$. The gap functions are given by $%
\delta\rho_s(\Delta,T)=\frac{2\rho_s(0)}{k_BT}\int^{\infty}_0 f(\epsilon,T)%
\cdot[1-f(\epsilon,T)]d\epsilon$, where $f(\epsilon,T)=(1+\exp{\sqrt{{%
\epsilon}^2+\Delta^2(T)}/k_BT})^{-1}$ is the Fermi distribution function.
Here, $k_B$ is the Boltzmann constant and $\Delta_i(T)$ were taken to follow
the universal BCS $T$-dependence.

Indeed the two-gap model can describe the overall superfluid density $\rho
_{s}(T)$ of Y$_{2}$C$_{3}$ as shown by the dotted line in Fig. 2,
particularly in the high temperature region near $T_{c}$. Such a fitting
gives the following parameters: $2\Delta _{1}/k_{B}T_{c}=4.9$, $2\Delta
_{2}/k_{B}T_{c}=1.1$ and $x=0.86$, which are in good agreement with those in
NMR \cite{Harada} and $\mu $SR measurements \cite{Kuroiwa}. However,
significant deviation of $\rho _{s}(T)$ from the two-gap model is observed
at temperatures below $0.4T_{c}$ (see inset (a) in Fig. 2). Our data clearly
show $\rho _{s}(T)\sim T$ at the lowest temperatures, providing strong
evidence for the occurrence of line nodes at low temperatures. Such behavior
can not be described by a simple two-gap BCS model. It is noted that our
reanalysis of the NMR data reported in Ref.\cite{Harada} suggests $%
1/T_{1}\sim T^{3}$ at $T<3$K (see inset (b) in Fig. 2), consistent
with our TDO measurements and further supporting the existence of
line nodes rather than a fully opened gap in Y$_{2}$C$_{3}$.
However, accurate measurements of the muon relaxation rate and/or
the nuclear spin-lattice relaxation rate at lower temperatures are
needed.

The nodal superconducting gap structure, together with a weak
Hebel-Slichter coherence peak \cite{Harada}, is rarely observed in a
simple metallic superconductor which shows no evidence of strong
electronic correlations and magnetic fluctuations. As we know, the
following three possibilities may lead to a nodal gap structure
within the phonon pairing mechanism. First, exotic superconductivity
with a nodal gap structure may appear in a multi-band system in
which the different sheets of Fermi surface are connected via some
"necks" \cite{Agterburg}. However, in this case the superconducting
transition is usually accompanied by some kind of magnetic order.
Indeed, the band structure calculations indicate that the Fermi
surface of Y$_{2}$C$_{3}$ consists of three hole bands and one
electron band, arising mainly from the Y$-4d$ and C$-2p\pi ^{\ast }$
hybridization \cite{Nishikayama}. The differences in the density of
states and Fermi velocities between hole and electron bands might
lead to two superconducting gaps opening in different parts of the
Fermi surface. However, there is no evidence of any magnetic order
in Y$_{2}$C$_{3}$ and, therefore, one may exclude such a scenario
here. Recently, Fujimoto studied the case of a non-centrosymmetric
superconductor with a weak ASOC and found that a field-induced pair
correlation between the different spin-orbit split bands might yield
a point-like anisotropic gap \cite{Fujimoto76}; this hardly
describes our experimental findings at zero field. As an
alternative, we argue that the ASOC effect arising from the broken
inversion symmetry might mix the spin-singlet and spin-triplet
pairing states, leading to the existence of line nodes in the
destructive gap as we previously observed in
Li$_{2}$(Pd$_{1-x}$Pt$_{x}$)$_{3}$B systems \cite{Yuan06,Yuan08}. In
the non-centrosymmetric superconductor Y$_{2}$C$_{3}$, the band
splitting due to ASOC effect is compatible with the superconducting
gap \cite{Nishikayama}, a situation similar to that of $x=0.3$ in
Li$_{2}$(Pd$_{1-x}$Pt$_{x}$)$_{3}$B \cite{Yuan08}. In this case, the
contribution of the spin-triplet component might be compatible with
or even slightly larger than that of the spin-singlet component,
resulting in an anisotropic gap with accidental nodes at a small
fraction of the Fermi surface which might then lead to a weak linear
temperature dependence in the penetration depth $\lambda (T)$ at the
lowest temperatures.

\begin{figure}[b]
\centering
\includegraphics[width=7cm]{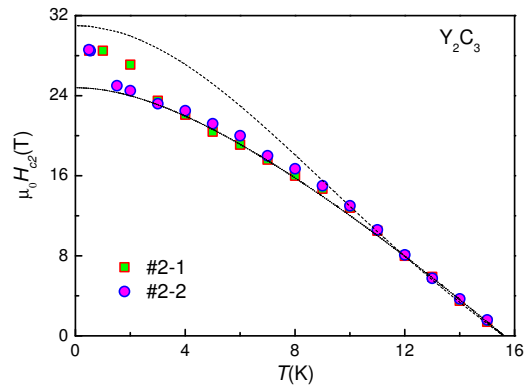}
\caption{(Color online) The upper critical field versus temperature for two
samples of Y$_2$C$_3$: \#2-1 and \#2-2. The dotted line and the dashed line
are fits to the weak coupling Werthamer-Helfand-Hohenberg (WHH) method and
the Ginzburg-Landau (GL) theory, respectively.}
\label{fig3}
\end{figure}

The upper critical field $\mu _{0}H_{c2}(T_{c})$ determined by a
similar TDO experiment in a pulsed magnetic field further supports
our argument mentioned above. In Fig. 3 we plot the temperature
dependence of $\mu _{0}H_{c2}(T_{c})$ down to 0.5K for the same
samples (\#2-1 and \#2-2). The following two points can be derived
from Fig. 3: First, the upper critical field shows very unusual
temperature dependence, $\mu _{0}H_{c2}(T_{c})$ increasing linearly
with decreasing temperature near $T_{c}$ but showing a weak upturn
at low temperatures where $\rho _{s}\sim T$. Such temperature
dependence of $\mu _{0}H_{c2}(T_{c})$ cannot be described either by
the weak coupling Werthamer-Helfand-Hohenberg (WHH) method
\cite{WHH} or the Ginzburg-Landau (GL) theory ($\mu
_{0}H_{c2}(T)=\mu _{0}H_{c2}(0)[1-(T/T_{c})^{2}]/[1+(T/T_{c})^{2}]$)
as shown in Fig. 3. Second, our direct measurements give a rather
high upper critical field of $\mu
_{0}H_{c2}(0)\simeq 29$ T, which is higher than the orbitally limited field (%
$\mu _{0}H_{c2}^{orb}(0)=-0.69T_{c}(d\mu _{0}H_{c2}/dT)_{T_{c}}=24.5$T) and
compatible with the paramagnetic limiting field ($\mu _{0}H_{c2}^{P}(0)=(1.86
$ T/K$)T_{c}=28.8$ T) derived for a weak BCS superconductor. Such an
enhancement of $\mu _{0}H_{c2}(0)$ was previously observed in the
non-centrosymmetric superconductors CePt$_{3}$Si, CeRhSi$_{3}$ and CeIrSi$%
_{3}$ \cite{Bauer04,Kimura98,Settai08}, presumably attributed to the
contribution of the spin-triplet component in the mixed pairing state \cite%
{Frigeri,Fujimoto9}.


In summary, we have investigated the penetration depth $\lambda (T)$ and the
upper critical field $\mu _{0}H_{c2}(T_{c})$ of the non-centrosymmetric
superconductor Y$_{2}$C$_{3}$ using a TDO-based resonant oscillator. We
found that a two-gap model can well describe the superfluid density $\rho
_{s}(T)$ at temperatures near $T_{c}$, but is clearly violated at low
temperatures. The frequency shift $\Delta f(T)$ and, therefore, the
corresponding penetration depth $\Delta \lambda (T)$ and superfluid density $%
\rho _{s}(T)$ show a weak linear temperature dependence at the lowest
temperatures, indicative of the existence of line nodes in the
superconducting energy gap. Together with the observation of an enhanced
upper critical field ($\mu _{0}H_{c2}(0)\simeq $ 29 T), we argue that these
nontrivial superconducting properties might be attributed to the absence of
an inversion symmetry in Y$_{2}$C$_{3}$, in which the ASOC splits the
electronic bands, mixing the spin-singlet and spin-triplet pairing states
and, therefore, leading to the existence of line nodes. Our findings
indicate that Y$_{2}$C$_{3}$ might present another important example to
study the effect of ASOC on superconductivity.


We acknowledge the helpful discussion with D. F. Agterburg, S. K.
Yip and M. Sigrist. This work was supported by Natural Science
Foundation of China, the National Basic Research Program of China,
the PCSIRT of the Ministry of Education of China, Zhejiang
Provincial Natural Science Foundation of China and the Fundamental
Research Funds for the Central Universities. Work at NHMFL-LANL is
performed under the auspices of the National Science Foundation,
Department of Energy and State of Florida. J.A. was partially
supported by ''High-Tech Research Center Project'' for Private
Universities and Grant-in-Aid for Scientific Research from Ministry
of Education, Culture, Sports, Science and Technology, Japan.

\end{document}